\documentclass{svjour3} 
\usepackage{graphics}
\usepackage{amsmath,amssymb}
\usepackage{natbib}
\usepackage{bm}

\smartqed
\journalname{CeMDA}

\begin{document}
\sloppy

\title{Lie-series for orbital elements}
\subtitle{II. The spatial case}

\author{Andr\'as P\'al}

\institute{%
A. P\'al 
	\at Konkoly Observatory of the MTA Research Centre for Astronomy and Earth Sciences, Budapest, Hungary and 
	\at Department of Astronomy, Lor\'and E\"otv\"os University, Budapest, Hungary \\    
E-mail: apal@szofi.net}

\maketitle

\begin{abstract}
If one has to attain high accuracy over long timescales during the
numerical computation of the $N$-body problem, the method called 
Lie-integration is one of the most effective algorithms. 
In this paper we present a set of recurrence relations with which the 
coefficients needed by the Lie-integration of the orbital elements related 
to the spatial $N$-body problem can be derived up to arbitrary order.
Similarly to the planar case, these formulae yields identically
zero series in the case of no perturbations. In addition, the derivation 
of the formulae has two stages, analogously to the planar problem. 
Namely, the formulae are obtained to the first order, and then, 
higher order relations are expanded by involving directly the multilinear 
and fractional properties of the Lie-operator. 
\end{abstract}

\section{Introduction}
\label{sec:introduction}

In terms of effectiveness, the method of Lie-integration is one 
of the most competitive algorithms for numerical computation 
of gravitational $N$-body dynamics. Unlike the ``classical'' ways
for numerical integration, this method computes the Taylor-coefficients
of the solution \citep[see][]{grobner1967}. Hence, the integration itself is 
relatively straightforward once these coefficients are known. The derivation 
of the Taylor-coefficients for a particular $\dot x_i=f_i(x_1,\dots,x_N)$ 
ordinary differential equation is based on the so-called Lie-operator.
Recalling the basics of this method, we define this operator as 
\begin{equation}
L:=\sum\limits_{i=1}^N f_i\frac{\partial}{\partial x_i},
\end{equation}
and by involving this definition, an advancement by $\Delta t$ 
of the ordinary differential equation can be written as
\begin{equation}
x_i(t+\Delta t)=\exp(\Delta t L)x_i(t)=\sum\limits_{k=0}^{\infty}\frac{\Delta t^k}{k!}L^kx_i(t).
\end{equation}
The numerical method called Lie-integration is the finite approximation of
the above equation for exponential expansion 
\citep[up to a certain order which can either be fixed or be 
adaptively varied, see also Sec.~3.1 in][]{pal2010}.
In order to effectively obtain these coefficients, recurrence formulae can 
be applied for the Cartesian coordinates of the orbiting bodies 
which are directly bootstrapped with the initial conditions.
Such formulae are known for the gravitational $N$-body problem
\citep{hanslmeier1984,pal2007}. A similar kind of 
relation has been obtained for the restricted 
three-body problem \citep{delva1984}, and 
relativistic and non-gravitational effects (such as Yarkovsky force)
can be included as well \citep{bancelin2012}.
In addition, semi-analytic calculations can also be performed 
to obtain parametric derivatives of observables with respect
to orbital elements \citep{pal2010}. 

In this paper we present such recurrence formulae for the orbital
elements in the case of spatial gravitational $N$-body problem. 
Recently, the relations for planar orbital elements have been
derived \citep{pal2014}. Therefore, our goal now is to extend these
relations to the third dimension by including the orbital elements
\emph{related} to the orbital inclination and ascending node. It should be
noted, however, that the relations are not obtained for the longitude of
ascending node directly, since it is meaningless in the $i\to 0$ limit. 

In the following section, Sec.~\ref{sec:lienbody}, we describe the 
problem itself and the recurrence relations for the Cartesian
coordinates and velocities. The discussion
of the spatial problem is split into three 
parts. Sec.~\ref{sec:lieangularmomentum} details the angular momentum vector
and the related orbital orientation. The next part, 
Sec.~\ref{sec:lieeccentricity} shows how the orbital eccentricity
can be treated in the spatial problem. The set of relations is ended with
the mean longitude (Sec.~\ref{sec:liemeanlongitude}). 
In Sec.~\ref{sec:higherorderderivatives} we demonstrate how higher order
derivatives are obtained. 
Our conclusions are summarized in Sec.~\ref{sec:summary}.

\section{The $N$-body problem}
\label{sec:lienbody}

If we consider Cartesian coordinates and velocities, the recurrence
relations for the spatial gravitational $N$-body problem 
have the same structure as in the planar case. Similarly to \cite{pal2014},
let us fix one of the bodies (e.g. the Sun in the case of the Solar System)
at the center and this body is orbited by $N$ additional ones, indexed
by $1\le i \le N$. In total, we deal with $1+N$ bodies, having a mass of
$M$ and $m_i$, respectively. If we denote the coordinates
and velocities of the $i$th body by $(x_i,y_i,z_i)$ and 
$(\dot x_i,\dot y_i,\dot z_i)$, we can define the central and mutual
distances $\rho_i$ and $\rho_{ij}$ as $\rho_i^2=x_i^2+y_i^2+z_i^2$
and $\rho_{ij}^2=(x_i-x_j)^2+(y_i-y_j)^2+(z_i-z_j)^2$, the
inverse cubic distances $\phi_i=\rho_i^{-3}$ and $\phi_{ij}=\rho_{ij}^{-3}$
and the standard gravitational parameters $\mu_i=G(M+m_i)$.
The quantities $\Lambda_i=x_i\dot x_i+y_i\dot y_i+z_i\dot z_i$,
and $\Lambda_{ij}=(x_i-x_j)(\dot x_i-\dot x_j)+(y_i-y_j)(\dot y_i-\dot y_j)+(z_i-z_j)(\dot z_i-\dot z_j)$
are also employed in the series of recurrence relations. 
With these quantities, the recurrence relations for the $x_i$ coordinates
and $\dot x_i$ velocities can be written as
\begin{eqnarray}
L^{n+1}x_i & = & L^n\dot x_i, \\
L^{n+1}\dot x_i & = & -\mu_i\sum\limits_{k=0}^n\binom{n}{k}L^k\phi_iL^{n-k}x_i \nonumber \\
& & -		\sum\limits_{j\ne i}Gm_j\sum\limits_{k=0}^n\binom{n}{k}\left[L^k\phi_{ij}L^{n-k}(x_i-x_j)+L^k\phi_jL^{n-k}x_j\right], 
\end{eqnarray}
while the relations for $y_i$ and $z_i$ also have the same structure. 
The relations for the reciprocal cubic distances can be computed in a similar
manner as it is done in the planar case, for instance, using 
Eqs.~(3)--(6) from \cite{pal2014}. Once the recurrence relations are
obtained and evaluated with the appropriate initial conditions,
temporal evolution can be computed with the finite approximation of
\begin{equation}
x_i(t+\Delta t)=\exp\left(\Delta t L\right)x_i(t)=\sum\limits_{k=0}^{\infty}\frac{(\Delta t)^k}{k!}L^kx_i(t)\approx\sum\limits_{k=0}^{k_{\rm max}}\frac{(\Delta t)^k}{k!}L^kx_i(t).
\end{equation}
Here the summation limit $k_{\rm max}$ refers to the maximum integration order.
Of course, this calculation is performed not only for the $x_i$ coordinates
but for all of the Cartesian coordinates and velocities.

\section{The angular momentum and the orientation of the orbit}
\label{sec:lieangularmomentum}

In the following, we detail the computations and relations comprehending
the orbital angular momentum and the orientation of the orbit. 

\subsection{Angular momentum}
\label{sec:angularmomentum}

In the case of the planar problem, the angular momentum is a 
pseudoscalar since it is the Hodge-dual of a skew-symmetric 
tensor of rank 2. In the spatial case, the angular momentum is 
still a skew-symmetric tensor of rank 2, hence it will have a 3 component 
dual in a form of a pseudovector. For the $i$th body, let us denote
these 3 components by $C_{xi}$, $C_{yi}$ and $C_{zi}$, respectively.
These are computed as 
\begin{eqnarray}
C_{xi} & = & y_i\dot z_i-z_i\dot y_i, \\
C_{yi} & = & z_i\dot x_i-x_i\dot z_i, \\
C_{zi} & = & x_i\dot y_i-y_i\dot x_i.
\end{eqnarray}
The first order Lie-derivatives of these pseudovector components
can similarly be computed like the pseudoscalar angular momentum in the
planar case, viz.
\begin{eqnarray}
LC_{xi} & = & \sum\limits_{j\ne i}Gm_j\hat\phi_{ij}S^{[x]}_{ij}\label{eq:cxlie}, \\
LC_{yi} & = & \sum\limits_{j\ne i}Gm_j\hat\phi_{ij}S^{[y]}_{ij}\label{eq:cylie}, \\
LC_{zi} & = & \sum\limits_{j\ne i}Gm_j\hat\phi_{ij}S^{[z]}_{ij}\label{eq:czlie},
\end{eqnarray}
where $S^{[x]}_{ij}$, $S^{[y]}_{ij}$ and $S^{[z]}_{ij}$ are defined as
\begin{eqnarray}
S^{[x]}_{ij} & = & y_iz_j-z_iy_j, \\
S^{[y]}_{ij} & = & z_ix_j-x_iz_j, \\
S^{[z]}_{ij} & = & x_iy_j-y_ix_j,
\end{eqnarray}
and $\hat\phi_{ij}=\phi_{ij}-\phi_j$. 
In order to compute the magnitude of the angular momentum vector, $C_i$, we 
can employ two approaches, as well. First, using the fact that $C_i^2$
is the sum of squares of the pseudovector components $C_{xi}$, $C_{yi}$ and $C_{zi}$,
we can write 
\begin{equation}
\frac{1}{2}L\left(C_i^2\right)=C_iLC_i=C_{xi}LC_{xi}+C_{yi}LC_{yi}+C_{zi}LC_{zi}.
\end{equation}
The second alternative is to exploit Lagrange's identity for cross products, namely 
\begin{equation}
\frac{1}{2}C_i^2=
\frac{1}{2}\mathbf{C}_i\cdot\mathbf{C}_i=
\frac{1}{2}(\mathbf{r}_i\times\dot{\mathbf{r}}_i)\cdot(\mathbf{r}_i\times\dot{\mathbf{r}}_i)=
\frac{1}{2}\mathbf{r}_i^2\dot{\mathbf{r}}_i^2-\frac{1}{2}(\mathbf{r}_i\cdot\dot{\mathbf{r}}_i)^2=
\frac{1}{2}\rho_i^2U_i^2-\frac{1}{2}\Lambda_i^2,\label{eq:csquare}
\end{equation}
where $U_i^2=\dot x_i^2+\dot y_i^2+\dot z_i^2$. Here, $\rho_i^2U_i^2$ 
can be written as $2\mu_i\rho_i-H_i\rho_i^2$ where $H_i$ is twice the 
negative specific energy, $H_i=2\mu_i/\rho_i-U_i^2$. Since both $H_i$ and
$\Lambda_i$ are scalars, the planar and spatial forms of the first 
Lie-derivatives are going to be the same:
\begin{eqnarray}
LH_i & = & 2\sum\limits_{j\ne i}Gm_j\left[\phi_{ij}\Lambda_i-\hat\phi_{ij}\hat\Lambda_{ji}\right], \\
L\Lambda_i & = & \left(U_i^2-\frac{\mu_i}{\rho_i}\right)+\sum\limits_{j\ne i}Gm_j\left[\hat\phi_{ij}R_{ij}-\phi_{ij}\rho_i^2\right].
\end{eqnarray}
Here $R_{ij}=x_ix_j+y_iy_j+z_iz_j$ and 
$\hat\Lambda_{ji}=x_j\dot x_i+y_j\dot y_i+z_j\dot z_i$
\citep[see also][]{pal2014}.
Using the relation $\frac{1}{2}L(\rho_i^2)=\Lambda_i$, 
the above two equations and Eq.~(\ref{eq:csquare}), it can be seen that
\begin{equation}
\frac{1}{2}L\left(C_i^2\right)=\sum\limits_{j\ne i}Gm_j\hat\phi_{ij}\left[\rho_i^2\hat\Lambda_{ji}-\Lambda_iR_{ij}\right].\label{eq:cliescalar}
\end{equation}
We should emphasize here that although $|C_{zi}|$ is equal to $C_i$ in 
the planar limit\footnote{When $z_i\to 0$ \emph{and} $\dot z_i\to 0$ 
for all $1\le i\le N$.}, it does not mean that expressions valid in the 
planar case could automatically be extended into the spatial form \emph{if}
such expressions are functions of pseudoscalars. In the calculations
presented in \cite{pal2014}, such differences were tacitly ignored,
therefore one should examine the individual terms before applying these in the
third dimension. In fact, $C_i=\sqrt{C_i^2}$ is a scalar (hence Eq.~\ref{eq:cliescalar}
is valid in both the planar and spatial cases), but $C_{zi}$ is not --
despite the validity of Eq.~(\ref{eq:czlie}) for the angular momentum
in the planar case. 

\subsection{The orientation of the orbit}
\label{sec:orientation}

Using the well known relations for the longitude of the ascending node $\Omega$
and the inclination $i$, one can compute these by knowing the components
of the angular momentum pseudovector:
\begin{eqnarray}
\sin i_i\cos\Omega_i 	& = & -\frac{C_{yi}}{C_i}, \label{eq:sinicosomega} \\
\sin i_i\sin\Omega_i 	& = & +\frac{C_{xi}}{C_i}, \label{eq:sinisinomega}\\
\cos i_i		& = & \frac{C_{zi}}{C_i}.
\end{eqnarray}
We note that in the case of small inclinations, the longitude of ascending node
is not so well constrained, so in order to avoid roundoff errors or
parametric singularities, it is easier to use the Lagrangian orbital elements
$\sin i_i\cos\Omega_i$ and $\sin i_i\sin\Omega_i$ instead of the angles. 
Due to the simple relations between the Lagrangian orbital elements and
the components of the angular momentum pseudovector, it is also sufficient
to deal purely with the $C_{xi}$, $C_{yi}$ and $C_{zi}$ terms. 

\subsection{Lie-series for fractions}
\label{sec:liefractions}

In the above relations for the Lagrangian ascending node and inclination,
fractions appear for quantities whose Lie-series are known. Although
recurrence relations for such fractions can be computed in two steps
(first by computing the denominator's reciprocal, then multiply it 
using the Leibniz' product rule with the numerator), it can be performed
in a single step. Let us have two quantities, $A$ and $B$ for which
the relations are known up to the order $n$. It can be shown by 
mathematical induction that the $n$th Lie-derivative of $A/B=AB^{-1}$
can be written as a function of the Lie-derivatives of $A$, $B$ up
to the order $n$ and $AB^{-1}$ up to the order $n-1$:
\begin{equation}
L^{n}(AB^{-1})=(L^nA)B^{-1}-B^{-1}\sum\limits_{k=1}^{n}\binom{n}{k}L^{n-k}(AB^{-1})L^{k}B.\label{eq:fractions}
\end{equation}
Employing this relation reduces the number of auxiliary quantities that 
would otherwise have to be introduced for the computation of (more 
complex) recurrence relations. 

\section{Eccentricity and related quantities}
\label{sec:lieeccentricity}

In the spatial case, the longitude of pericenter, $\varpi$ is defined
as the sum of longitude of ascending node $\Omega_i$ and the argument
of pericenter, $\omega_i$, namely $\varpi_i=\Omega_i+\omega_i$. This definition
yields the continuity of the longitude of pericenter in the planar
limit of $i_i\to 0$ when both $\Omega_i$ and $\omega_i$ are meaningless.
Once $\varpi_i$ is obtained, the Lagrangian orbital elements $k_i$ 
and $h_i$ are defined accordingly, i.e.
\begin{equation}
\binom{k_i}{h_i}=e_i\binom{\cos\varpi_i}{\sin\varpi_i}.
\end{equation}
It can also be deduced that if the $i$th orbit is rotated around the line
of its nodes into the reference plane then $\varpi_i$, and hence 
$k_i$ and $h_i$ are not altered. The aforementioned 
rotation depends only on the components of the angular momentum vector.
Hence, we can write the related rotation matrix as the function of the 
$C_{xi}$, $C_{yi}$ and $C_{zi}$ components as
\begin{equation}
\begin{pmatrix}
1-\dfrac{C_{xi}^2}{C_i^2+C_iC_{zi}}\hspace*{4mm} & \dfrac{-C_{xi}C_{yi}}{C_i^2+C_iC_{zi}}& -\dfrac{C_{xi}}{C_i} \\[5mm]
-\dfrac{C_{xi}C_{yi}}{C_i^2+C_iC_{zi}} & 1-\dfrac{C_{yi}^2}{C_i^2+C_iC_{zi}}\hspace*{4mm}  & -\dfrac{C_{yi}}{C_i} \\[5mm]
\dfrac{C_{xi}}{C_i} & \dfrac{C_{yi}}{C_i} & \dfrac{C_{zi}}{C_i}
\end{pmatrix}
\end{equation}
For instance, the coordinate $x_i$ located in the $i$th 
orbital plane is transformed into:
\begin{equation}
x'_i = x_i - \frac{C_{xi}^2}{C_i^2+C_iC_{zi}}x_i - \frac{C_{xi}C_{yi}}{C_i^2+C_iC_{zi}}y_i-\frac{C_{xi}}{C_i}z_i.
\end{equation}
By exploiting the fact that $C_{xi}x_i+C_{yi}y_i+C_{zi}z_i=0$, the above
equation can greatly be simplified:
\begin{equation}
x'_i = x_i - \frac{C_{xi}z_i}{C_i+C_{zi}}. \label{eq:xirotated}
\end{equation}
The similar structure can be used for the $y_i$ coordinate and 
the velocity components are also transformed similarly
since $C_{xi}\dot x_i+C_{yi}\dot y_i+C_{zi}\dot z_i$ is also $0$. 
Due to the previously noted invariance
of the $k_i$ and $h_i$ elements, these can be computed as 
\begin{equation}
\binom{k_i}{h_i}=\frac{C_i}{\mu_i}\binom{+\dot y'_i}{-\dot x'_i}-\frac{1}{\rho_i}\binom{x'_i}{y'_i}.
\end{equation}
If we substitute Eq.~\ref{eq:xirotated} (and the similar relations
for $y_i$, $\dot x_i$ and $\dot y_i$) into the above equation we get
\begin{equation}
\binom{k_i}{h_i}=\frac{C_i}{\mu_i}\left[\binom{+\dot y_i}{-\dot x_i}-\binom{+p_{yi}}{-p_{xi}}\dot z_i\right]
-\frac{1}{\rho_i}\left[\binom{x_i}{y_i}-\binom{p_{xi}}{p_{yi}}z_i\right],
\end{equation}
where we defined
\begin{equation}
\binom{p_{xi}}{p_{yi}}=\frac{1}{C_i+C_{zi}}\binom{C_{xi}}{C_{yi}}.
\end{equation}
We note that these $p_{xi}$ and $p_{yi}$ quantities are also integrals of motion
and can be computed purely from the inclination and longitude of ascending node
(but not as simple as in Eqs.~\ref{eq:sinicosomega} or \ref{eq:sinisinomega}).
Let us also define the quantities $a_{xi}$, $a_{yi}$, $a_{zi}$ as 
\begin{eqnarray}
a_{xi} & = & \sum\limits_{i\ne j} Gm_j [\hat\phi_{ij}x_j-\phi_{ij}x_i]\label{eq:axi}, \\
a_{yi} & = & \sum\limits_{i\ne j} Gm_j [\hat\phi_{ij}y_j-\phi_{ij}y_i]\label{eq:ayi}, \\
a_{zi} & = & \sum\limits_{i\ne j} Gm_j [\hat\phi_{ij}z_j-\phi_{ij}z_i]\label{eq:azi}.
\end{eqnarray}
Using the previously introduced variables, we can compute the first
order Lie-derivatives of $k_i$ and $h_i$ as 
\begin{eqnarray}
Lk_i & = & +\left[\frac{LC_i}{\mu_i}\dot y_i+\frac{C_i}{\mu_i}a_{yi}\right]
	-p_y\left[\frac{LC_i}{\mu_i}\dot z_i-\frac{C_i}{\mu_i}a_{zi}\right]
	-\left[+\frac{C_i}{\mu_i}Lp_{yi}\dot z_i-\frac{z_i}{\rho_i}Lp_{xi}\right], \label{eq:liek} \\
Lh_i & = & -\left[\frac{LC_i}{\mu_i}\dot x_i+\frac{C_i}{\mu_i}a_{xi}\right]
	+p_x\left[\frac{LC_i}{\mu_i}\dot z_i-\frac{C_i}{\mu_i}a_{zi}\right]
	-\left[-\frac{C_i}{\mu_i}Lp_{xi}\dot z_i-\frac{z_i}{\rho_i}Lp_{yi}\right]. \label{eq:lieh}
\end{eqnarray}
Here, the first order derivatives $Lp_{xi}$ and $Lp_{yi}$ can be computed as 
\begin{equation}
\binom{Lp_{xi}}{Lp_{yi}}=\frac{1}{C_i+C_{zi}}\left[\binom{LC_{xi}}{LC_yi}-(LC_i+LC_{zi})\binom{p_{xi}}{p_{yi}}\right].\label{eq:lpxyi}
\end{equation}
The derivation of the above equations is similar to the steps
performed in \cite{pal2014}. The above two equations for $Lk_i$ and $Lh_i$
are clearly zero if mutual perturbations are omitted since then
$LC_i$, $a_{xi}$, $a_{yi}$, $a_{zi}$, $Lp_{xi}$ and $Lp_{yi}$ are zero.

As an alternative, one can compute the Lie-derivatives of the 
Laplace-Runge-Lenz vector. In the spatial case, this vector is defined as 
\begin{equation}
\mathbf{e}_i=\frac{1}{\mu_i}(\dot{\mathbf{r}}_i\times\mathbf{C}_i)-\frac{\mathbf{r}_i}{\rho_i},
\end{equation}
while all of its components,
\begin{eqnarray}
e_{xi} & = & \frac{1}{\mu_i}(C_{zi}\dot y_i-C_{yi}\dot z_i)-\frac{x_i}{\rho_i}, \\
e_{yi} & = & \frac{1}{\mu_i}(C_{xi}\dot z_i-C_{zi}\dot x_i)-\frac{y_i}{\rho_i}, \\
e_{zi} & = & \frac{1}{\mu_i}(C_{yi}\dot x_i-C_{xi}\dot y_i)-\frac{z_i}{\rho_i}
\end{eqnarray}
are integrals of motion. The Lie-derivatives of each of these components have 
the same structure and can be obtained in a similar manner to the planar case. 
The first order Lie-derivatives of the $(e_{xi},e_{yi},e_{zi})$ 
components are
\begin{eqnarray}
Le_{xi} & = & \frac{1}{\mu_i}\left[LC_{zi}\dot y_i+C_{zi}a_{yi}-LC_{yi}\dot z_i-C_{yi}a_{zi}\right], \\
Le_{yi} & = & \frac{1}{\mu_i}\left[LC_{xi}\dot z_i+C_{xi}a_{zi}-LC_{zi}\dot x_i-C_{zi}a_{xi}\right], \\
Le_{zi} & = & \frac{1}{\mu_i}\left[LC_{yi}\dot x_i+C_{yi}a_{xi}-LC_{xi}\dot y_i-C_{xi}a_{yi}\right].
\end{eqnarray}
In a practical implementation, one could choose whether to compute
the Lagrangian orbital elements $k_i$, $h_i$ or the components of 
the vector $\mathbf{e}_i$. Due to the constraint 
\begin{equation}
\mathbf{C}_i\cdot\mathbf{e}_i=C_{xi}e_{xi}+C_{yi}e_{yi}+C_{zi}e_{zi}=0,
\end{equation}
these two sets of variables are equivalent. The first order
Lie-derivatives of both $(k_i,h_i)$ and $(e_{xi},e_{yi},e_{zi})$
are multilinear expressions of terms whose derivatives are known in advance. 
Therefore, higher order derivatives can be computed in a straightforward
manner: either using Eq.~(24) of \cite{pal2014} or by introducing
auxiliary variables and exploit the product rule for differentials. 

\section{Mean longitude}
\label{sec:liemeanlongitude}

In order to compute the Lie-derivatives of the mean longitude, we can employ
two different approaches. First, similarly to \cite{pal2014}, we write
a relatively complex equation for it and then take the full derivative.
Here we follow an alternative approach. First, let us 
write the mean longitude in the form of $\lambda_i = M_i + \varpi_i$, 
where $M_i$ is the mean anomaly\footnote{Note that the symbol $M$ 
represents the central mass while the symbols $M_i$ (with a single index)
denote the mean anomalies.} and $\varpi_i=\arg(k_i,h_i)$ is the 
longitude of pericenter. Then take the first order Lie-derivatives of both, 
coadd them in the hope that in the circular limit, the sum $LM_i + L\varpi_i$ 
would not be meaningless. Finally, we use this first order derivative in 
order to obtain the recurrence relations for higher order Lie-derivatives.

According to Kepler's equation, the mean anomaly is computed as 
$M_i=E_i-e_i\sin E_i$ where the eccentric anomaly is written in the form of
\begin{equation}
E_i=\arg(e_i\cos E_i,e_i\sin E_i).
\end{equation}
Although $E_i$ is still meaningless
in the $e_i\to0$ limit, the terms $e_i\cos E_i$ and $e\sin E_i$ can be 
computed using the basic relations of two-body kinematics even in the
circular case: 
\begin{eqnarray}
e_i\cos E_i & = & 1-\frac{\rho_i}{a_i} = 1-\frac{\rho_iH_i}{\mu_i}, \label{eq:ecose}\\
e_i\sin E_i & = & \frac{\Lambda_i J_i}{C_i}, \label{eq:esine}
\end{eqnarray}
where $J_i:=\sqrt{1-e_i^2}$ \citep[similarly to the definition used in][]{pal2014}.
Then, the first order Lie-derivative of $M_i$ is going to be
\begin{eqnarray}
LM_i & = & L\arg(e_i\cos E_i,e_i\sin E_i)-L(e_i\sin E_i)=  \nonumber \\
& = & \frac{e_i\cos E_iL(e_i\sin E_i)-e_i\sin E_iL(e_i\cos E_i)}{e_i^2}-L(e_i\sin E_i).
\end{eqnarray}
After multiplying by $e_i^2$ and substituting Eqs.~(\ref{eq:ecose}) 
and (\ref{eq:esine}), we get
\begin{equation}
e_i^2LM_i=\left(1-\frac{\rho_iH_i}{\mu_i}\right)L\left(\frac{\Lambda_iJ_i}{C_i}\right)+\frac{\Lambda_iJ_i}{C_i}L\left(\frac{\rho_iH_i}{\mu_i}\right)-(1-J_i^2)L\left(\frac{\Lambda_iJ_i}{C_i}\right).
\end{equation}
The expansion of the above equation yields the form
\begin{equation}
e_i^2LM_i=\mu_i^2\frac{J_i^3}{C_i^3}e_i^2+\frac{J_i^3}{C_i}\left(1-\frac{\rho_i\mu_i}{C_i^2}\right)L\Lambda_{pi}+\left(1+\frac{\rho_i\mu_i}{C_i^2}\right)\frac{J_i\Lambda_iC_i}{2\mu_i^2}LH_i. \label{eq:eeliemeananomaly}
\end{equation}
In this expansion, we use the quantity $L\Lambda_{pi}$ which is defined 
as follows. Since $\Lambda_i$ is not an integral of motion, we 
split $L\Lambda_i$ into two parts, viz.
\begin{equation}
L\Lambda_i=\left(U_i^2-\frac{\mu_i}{\rho_i}\right)+\sum\limits_{i\ne j}Gm_j\left[\hat\phi_{ij}R_{ij}-\phi_{ij}\rho_i^2\right]=\left(U_i^2-\frac{\mu_i}{\rho_i}\right)+L\Lambda_{pi}
\end{equation}
and then define $L\Lambda_{pi}$ accordingly. Eq.~(\ref{eq:eeliemeananomaly})
for the mean anomaly has three parts. The first one correspond to Kepler's 
third law after dividing by $e_i^2$. Despite the fact that in the non-perturbed
case, the other two parts are zero, in the perturbed case (when $L\Lambda_{pi}\ne 0$ 
or $LH_i\ne 0$), the multipliers are only $\mathcal{O}(e_i)$ functions,
not $\mathcal{O}(e_i^2)$ functions, 
therefore $LM_i$ is meaningless in the $e_0\to 0$ limit.

The first order Lie-derivative of the mean longitude can only be computed
if $e_i^2L\varpi_i$ is added to Eq.~(\ref{eq:eeliemeananomaly}). The
derivative of $\varpi_i$ is computed using the relation
\begin{equation}
e_i^2L\varpi_i=k_iLh_i-h_iLk_i. \label{eq:lievarpi}
\end{equation}
It can be shown that if we add Eq.~(\ref{eq:lievarpi}) to the equation
related to the mean anomaly (see Eq.~\ref{eq:eeliemeananomaly}), 
all of the $\mathcal{O}(e_i)$ terms cancel
and $L\lambda_i$ is continuous in the $e_i\to 0$ limit. Without going 
into the details, here we present the results of this computation.
Similarly to the planar case, $L\lambda_i$ is written into two parts:
the first part corresponds to Kepler's third law while the another
terms depend only on the mutual perturbations. Namely, 
\begin{eqnarray}
L\lambda_i & = & \frac{1}{\mu_i}H_i^{3/2} + 
	A_0\rho_i^2\sum\limits_{j\ne i}Gm_j\phi_{ij} +
	A_{\rm A}\sum\limits_{j\ne i}Gm_j\hat\phi_{ij}(C_{xi}x_j+C_{yi}y_j) +
	\nonumber\\
& & +	A_{\rm z}\sum\limits_{j\ne i}Gm_j\hat\phi_{ij}z_j + 
	A_{\rm P}\sum\limits_{j\ne i}Gm_j\hat\phi_{ij}R_{ij} +
	A_{\rm L}\sum\limits_{j\ne i}Gm_j\hat\phi_{ij}\hat\Lambda_{ji}\label{eq:llambdai}
\end{eqnarray}
The expressions for $A_0$, $A_{\rm A}$, $A_{\rm z}$, $A_{\rm P}$
and $A_{\rm L}$ are the following:
\begin{eqnarray}
A_0 		& = & \frac{1}{C_i}\left(\frac{g_i^{-2}-g_i^{-1}}{1+J_i}+2J_i\right), \label{eq:a0} \\
A_{\rm A}	& = & \frac{z_i}{(1+\cos i_i)^2C_i^2}, \label{eq:aa} \\
A_{\rm z}	& = & \frac{z_i}{(1+\cos i_i)C_i}, \label{eq:az} \\
A_{\rm P}	& = & 	\frac{1}{C_i}\left(\frac{J_i^2(g_i-1)}{1+J_i}-2\right)	
			+\frac{z_i\left[C_i^2\Lambda_i\dot z_i-\mu_i^2(2g_i^{-1}-J_i^2)z_i\right]}{C_i^5(1+\cos i_i)^2},
			\label{eq:ar}  \\
A_{\rm L}	& = & 	\frac{\Lambda_iC_i(1+g_i)}{\mu_i^2(1+J_i)}
			+\frac{z_i\left[-C_i^4g_i^2\dot z_i+\Lambda_i\mu_i^2z_i\right]}{C_i^3\mu_i^2(1+\cos i_i)^2},
			\label{eq:al} 
\end{eqnarray}
where the dimensionless quantity $g_i$ is defined as $g_i:=\mu_i\rho_iC_i^{-2}$.

One should note that the quantity $A_0$ equals to the quantity with
the same name used in Eq.~(49) of \cite{pal2014}. We should also warn the
reader that in the purely planar case, the expansion of $L\lambda_i$
involved the quantity $\hat C_{ji}:=x_j\dot y_i-y_j\dot x_i$. Since
this quantity behaves as a pseudoscalar in the purely planar case, it 
has no direct counterpart in the framework of the spatial problem. Hence,
in Eq.~(\ref{eq:llambdai}) we express $L\lambda_i$ as the function
of $\hat\Lambda_{ji}$ instead of such pseudoscalar-like quantities. Therefore,
the equivalence of Eq.~(\ref{eq:llambdai}) here and Eq.~(49) of \cite{pal2014}
is not obvious at the first glance in the limit of $z\to0$ and $\dot z\to0$.
Nevertheless, one could verify this equivalence by considering the relation
$\hat\Lambda_{ji}^2+\hat C_{ji}^2=\rho_j^2 U_i^2$ (where $\hat C_{ji}$ cannot 
even be defined in the spatial case). 

We should note that some of the $A_{\rm *}$ terms explicitly 
contain the third coordinate, $z_i$ and/or its derivative, $\dot z_i$. 
Therefore, in a perturbed system, the time derivative of the mean 
longitude is not a scalar and this is only invariant for a subgroup of
the group ${\rm SO}(3)$ of proper rotations. This subgroup is 
the ${\rm SO}(2)$ rotations around the $z\pm$ axis. On the contrary,
the expression for the derivative of mean anomaly, $LM_i$ 
(see Eq.~\ref{eq:eeliemeananomaly}) is a function of scalars.
Hence, $LM_i$ is invariant under arbitrary ${\rm SO}(3)$ transformations.

\section{Higher order derivatives}
\label{sec:higherorderderivatives}

Higher order Lie-derivatives can then almost automatically be derived since
all of the corresponding expressions contain multilinear, power
and fractional terms for which recurrence relations are known. 
The bilinear relation follows Leibniz' product rule, for fractions one
can use the derivation presented in Sec.~\ref{sec:liefractions} while
for powers, one can involve Eq.~(51) of \cite{pal2014}. In brief,
one can conclude that the Lie-derivatives of \emph{any} rational function can 
be computed once the Lie-derivatives of the terms appearing in the function 
are known in advance. 

Actually, higher order relations for the angular momentum 
based on Eqs.~(\ref{eq:cxlie})-(\ref{eq:czlie}) can be written as 
\begin{eqnarray}
L^{n+1}C_{xi} & = & \sum\limits_{j\ne i}Gm_j\sum\limits_{k=0}^n\binom{n}{k}L^{n-k}\hat\phi_{ij}L^{k}S^{[x]}_{ij}\label{eq:cxlien}, \\
L^{n+1}C_{yi} & = & \sum\limits_{j\ne i}Gm_j\sum\limits_{k=0}^n\binom{n}{k}L^{n-k}\hat\phi_{ij}L^{k}S^{[y]}_{ij}\label{eq:cylien}, \\
L^{n+1}C_{zi} & = & \sum\limits_{j\ne i}Gm_j\sum\limits_{k=0}^n\binom{n}{k}L^{n-k}\hat\phi_{ij}L^{k}S^{[z]}_{ij}\label{eq:czlien},
\end{eqnarray}
where the corresponding derivatives of $L^k\hat\phi_{ij}$ are known from earlier
works \citep{hanslmeier1984,pal2007,pal2014} while 
\begin{eqnarray}
L^nS^{[x]}_{ij} & = & \sum\limits_{k=0}^n\binom{n}{k}\left(L^{n-k}y_iL^kz_j-L^{n-k}z_iL^ky_j\right), \\
L^nS^{[y]}_{ij} & = & \sum\limits_{k=0}^n\binom{n}{k}\left(L^{n-k}z_iL^kx_j-L^{n-k}x_iL^kz_j\right), \\
L^nS^{[z]}_{ij} & = & \sum\limits_{k=0}^n\binom{n}{k}\left(L^{n-k}x_iL^ky_j-L^{n-k}y_iL^kx_j\right).
\end{eqnarray}

Higher order relations for the Lagrangian orbital elements $k$ and $h$ 
are obtained by expanding the bi- and trilinear terms of Eqs.~(\ref{eq:liek})
and (\ref{eq:lieh}). This expansion has the following substeps:
\begin{itemize}
\item First, the fraction $L^{n}\left[(C_i+C_{zi})^{-1}\right]$ is needed
to be computed, using the rule presented in Sec.~\ref{sec:liefractions}.
Here, the numerator is $1$ (with zero Lie-derivatives), so Eq.~(\ref{eq:fractions})
can further be simplified. Alternatively, Eq.~(51) of \cite{pal2014} can
be used considering the exponent of $p=-1$. 
\item Once $L^{n}\left[(C_i+C_{zi})^{-1}\right]$ is known, $L^{n+1}p_{xi}$ 
and $L^{n+1}p_{xi}$ are derived using the trilinear Leibniz' product rule
for Eq.~\ref{eq:lpxyi}. 
\item The higher order derivatives of the accelerations $a_{xi}$, $a_{yi}$
and $a_{zi}$ are obtained using Leibniz' rule for two multiplicands,
following Eqs.~(\ref{eq:axi})-(\ref{eq:azi}). 
\item Once these three above steps are done, all of the terms are
known appearing in Eqs.~(\ref{eq:liek}) and (\ref{eq:lieh}). Hence, 
the trilinear rule should be applied. 
\end{itemize}
In a practical implementation, a programmer needs to treat 
$\left[(C_i+C_{zi})^{-1}\right]$ as a separate variable and store it 
accordingly in conjunction with its higher order derivatives. 
In addition, a trilinear expansion can also be speeded up if a product
like $L^n(ABC)$ is expanded in two bilinear substeps, namely first one
compute $L^n(AB)$ in the usual manner then $L^n(ABC)$ is written as 
\begin{equation}
L^{n}(ABC)=\sum\limits_{k=0}^n\binom{n}{k}L^{n-k}(AB)L^kC.
\end{equation}
This kind of optimization reduces the number of operations from
$\mathcal{O}(n^2)$ to $\mathcal{O}(n)$, however, auxiliary variables and 
the respective arrays are needed to be introduced.

The higher order relations for $L^{n+1}\lambda_i$ can also be considered
similarly since the terms appearing in Eq.~(\ref{eq:llambdai}) are 
bi-, tri- or quadrilinear functions of the terms $A_{x}$ and 
quantities for which the recurrence relations have already been obtained.
The terms $A_0$, $A_{\rm A}$, $A_{\rm z}$, $A_{\rm P}$ and $A_{\rm L}$
are complex expressions, however, these are still \emph{rational functions}
of quantities for which the recurrence series are known. 

\section{Summary}
\label{sec:summary}

This paper completes the recurrence relations for the Lie-derivatives
of the osculating orbital elements in the case of the spatial 
$N$-body problem. These relations can be exploited to integrate directly 
the equations of motions that are parameterized via the orbital elements.
Qualitatively, the advantages and disadvantages of this approach
are the same what has been concluded for the planar problem. Namely,
evolving orbital elements instead of Cartesian components results in
larger stepsizes. On the other hand, the complex implementation
and the need of more computing power (for the actual evaluation 
a single step) could yield only marginal benefit.
An initial implementation for a demonstration and validation of
the formulae presented in this article can be downloaded from our 
web page\footnote{http://szofi.elte.hu/\~{ }apal/astro/orbitlie/}
as well as these codes are available upon request. They are also
included in the supplement appended to the electronic version of the paper.

Correspondingly to the planar case, coordinates and velocities do appear 
in the recurrence relations but in a form of purely auxiliary quantities.
Further studies can therefore focus on the elimination of the need of 
coordinates. This is particularly interesting in the case of 
mean longitude where the third direction is preferred. Such derivations
might significantly reduce the computing demands as well. 

\vspace*{1ex}

\noindent
\textbf{Acknowledgments.}
The author would like to thank A. L\'aszl\'o for his helpful comments about 
the tensor rank analysis. The author also thanks the anonymous referees for
their thorough reviews of the manuscript. The author thanks L\'aszl\'o Szabados 
for the careful proofreading. This work has been supported 
by the Hungarian Academy of Sciences via the grant LP2012-31. 
Additional support is received from the Hungarian OTKA grants 
K-109276 and K-104607. 

{}

\end{document}